\documentclass{ceab}   

\usepackage{epsfig}     
\usepackage{graphicx}   

\usepackage{ceabbib}     
\usepackage[T1]{fontenc}

\def\gore{ E.\,CHMIELEWSKA, M.\,TOMCZAK, T.\,MROZEK}

\begin{document}

\title{Simultaneous \textit{Yohkoh}/SXT and \textit{TRACE} observations of solar plasma ejections}

\author{\normalsize E.\,CHMIELEWSKA$^{1}$, M.\,TOMCZAK$^{1}$, T.\,MROZEK$^{1, 2}$ \vspace{2mm} \\
        \it $^{1}$Astronomical Institute, University of Wroc{\l }aw, \\
        \it  ul.\,Kopernika 11, PL-51-622 Wroc{\l }aw, Poland\\
        \it $^{2}$Space Research Center, Polish Academy of Sciences, Solar Physics Division,\\ 
        \it ul.\,Kopernika 11, PL-51-622 Wroc{\l }aw,  Poland}

\maketitle

\begin{abstract}
We investigated the relationship between hot ($\sim$10 MK) and warm ($\sim$1 MK) plasma ejections 
observed in the solar corona during some flares from April 1998 to December 2001. Based on the catalog of X-ray Plasma Ejections 
(XPEs) observed by the Soft X-ray Telescope onboard the \textit{Yohkoh} satellite, we have identified counterparts of XPEs 
in extreme-ultraviolet (EUV) images, derived by the \textit{TRACE} satellite. We found that the XPEs are always associated with 
the EUV ejections.
In most cases, ejections of warm plasma start earlier than ejections of hot plasma. The comprehensive analysis of kinematic and morphology evolution of the most interesting and well-observed events is also given.

\end{abstract}


\section{Introduction}
Over the past decade it has become clear that there are three main classes of solar eruption events: prominences, coronal mass ejections (CMEs), and X-ray plasma ejections (XPEs) released due to destabilisation of solar magnetic fields. Understanding the process that generates and drives these phenomena will give us insight into the basics solar eruptive events, and will allow us to predict them.

XPEs are rapid expulsions of hot magnetized plasma in the solar corona observed in X-rays.
They have been systematically observed since 1991 when the {\it Yohkoh} satellite began to operate (Tomczak \& Chmielewska, 2012).

XPEs are a strongly inhomogeneuos group of events, showing different morphological and kinematical properties.
Plasma temperatures in these phenomena can reach several tens of millions Kelivn (Ohyama \& Shibata, 1998, 2008; Tomczak, 2004, Tomczak \& Ronowicz, 2007).
XPEs have a wide speed range from several km/s to a few thousands km/s, but most of the events are slower than 200 km/s (Tomczak, 2003; Kim {\it et al.}, 2005b; Tomczak \& Ronowicz, 2007).

Observations clearly show that particular classes of solar eruptions are connected with each other on different levels of correlation. Thus, a statistical investigation enables us not only to better understand relationships between different solar phenomena but also gives us an  insight into the complex configuration of magnetic field in the solar atmosphere. 

Prominence-CME pairs display close association from 67\% (Gilbert {et al.}, 2000) to 72\% (Gopalswamy {\it et al.}, 2003). Worthy of note is the fact that different subclasses of prominences show different levels of these correlations (Gilbert {et al.}, 2000; Gopalswamy {\it et al.}, 2003).

The association between XPEs and CMEs also is close as well as different for different subclasses of XPEs with an overall correlation of 66\% (Tomczak \& Chmielewska, 2012), slightly smaller than an earlier value of 69\% (Kim {\it et al}, 2005a).  

XPEs and prominences seem to be correlated poorer than the XPE -- CME and prominence -- CME correlations. Slightly above 30\% of XPEs is correlated with prominences and the value varies for particular subclasses of XPEs from about 19\% to 47\% (Chmielewska \& Tomczak, 2012).
 
\section{Association of XPEs with the EUV ejections}
 In order to make progress in our research on the relationship between basic classes of solar eruptions, we have looked for counterparts of XPEs from the catalog (Tomczak \& Chmielewska, 2012) in EUV images. In our analysis we used the observations from two space-borne imaging instruments {\it Yohkoh}/SXT and {\it TRACE}. {\it Yohkoh}/SXT was able to observe plasma in the temperature range from 1 to 50 MK, whereas {\it TRACE} monitored the Sun in several wavelengths covering a temperature range from thousands of Kelvins up to 2 MK. 

For each event from the studied XPEs we searched for an EUV counterpart in two steps: we checked coordinates of an active region, which was observed by {\it TRACE}  around the start of an XPE $\pm$10 minutes. When these two telescopes observed the same active region, we compared images from SXT and {\it TRACE} carefully to confront the position of individual structures and we looked for signatures of plasma movements in the images taken by {\it TRACE}. In order to identify a solar eruption observed in the EUV it was necessery to correct {\it  TRACE} pointing information as described in Peter T. Gallagher's page (http://www.tcd.ie/Physics/people/Peter.Gallagher/trace-align/index.html).  

\begin{table}[!t]
\caption{Correlation between the EUV ejections and other solar eruptive phenomena}
\label{tab:tab1}
\begin{center}

\begin{tabular}{@{\hspace{0.5cm}}l@{\hspace{0.5cm}}l@{\hspace{0.5cm}}l@{\hspace{0.5cm}}l}
\hline
\noalign{\hrule height 1.1 pt}

 &Association &Association& Association \\
\vspace{0.08 cm}
&with XPEs& with CMEs&with EPs\\ [0.1 em]

\hline
\noalign{\hrule height 1.1 pt}
\vspace{0.08 cm}
& 58/65 (89.2\%) & 39/51 (76.5\%) & 7/58 (12.1\%)  \\ [0.1 em]

\hline

\end{tabular}
\end{center}
\end{table}
\begin{table}[t]
\caption{Time differences between the start of XPEs and the start of EUV ejections}
\label{tab:tab2}
\begin{center}

\begin{tabular}{@{\hspace{0.5cm}}l@{\hspace{1cm}}l}
\hline
\noalign{\hrule height 1.2pt}

 $t_{ XPE \; start}-t_{EUV\; start}$& Quantity \\

\hline
\noalign{\hrule height 1.2 pt}
$<\!-1$ min& 40/54 (74.1 \%)\\

$\pm1$ min& 6/54 (11.1 \%)  \\

$>1$ min& 8/54 (14.8 \%)  \\
\hline

\end{tabular}
\end{center}
\end{table}


The results of our statistical analysis are summarized in Table \ref{tab:tab1}. 
We found that slightly below 90\% of eruptions observed in EUV are correlated with the XPEs. Furthermore, based on the data from the LASCO coronagraphs ({\it SOHO} LASCO CME Catalog (Gopalswamy {\it et al.}, 2009)) and the list of active prominences and filaments from comprehensive reports of the {\it Solar-Geophysical Data}, we obtained that 39 from 51 warm ejections were associated with CMEs (76.5\%), and only 7 from 58 EUV ejections were correlated with prominences and filaments (12\%).

In Table \ref{tab:tab2}, we present the results of time coincidence between the EUV ejections and XPEs. In 40 events from 54 we obtained negative values greater than 1 minute, which means that warm plasma ejetions observed by {\it TRACE} start before the XPEs in most cases. About 11\% of investigated events occurred almost at the same time as XPEs (6/54). For almost 15\% of events the start of warm ejections is seen after the start of XPEs.

We also examined the relative initial locations of XPEs and EUV ejections. 
In our analysis, we considered only 40 pairs of warm ejection-XPE, for which
determination of the initial altitude was possible.
For these events we observed 20 EUV ejections located higher than XPEs, and the same number of XPEs was situated higher than the TRACE ejections.
These results may suggest that the relative location of  ejections of warm and hot plasma is dependent on the individual conditions and magnetic configurations in active regions.

\section{Classification scheme}

For a more detailed comparison between XPEs and EUV ejections, we adopted a classification scheme developed in the XPE Catalog (Tomczak \& Chmielewska, 2012). Events in this catalog were classified according to three criteria: morphological, kinematical, and recurrence. For each criterion two subclasses of events were defined: (first criterion) collimated/loop-like, (second criterion) confined/eruptive, (third criterion) single/recurrent. 

The morphological criterion involves the direction of the movement of XPEs in comparison with the direction of the local magnetic field. For collimated events, the direction is along
the pre-existing magnetic field lines, while in the case of loop-like events, the motion is across the pre-existing field lines (or strictly speaking, together with them).

For an XPE assignment into one of the kinematical subclasses
we have chosen the height increase rate above the chromosphere,
$\dot{h} $. A negative value, $\dot{h} < 0$, means confined; the
opposite case, $\dot{h} \ge  0$  means an eruptive event.
Events from the kinematical subclass named confined suggest the presence of a kind of magnetic
or gravitational confinement of X-ray plasma. For eruptive XPEs, an increasing velocity in
the radial direction in the field of view of the SXT allows us
to anticipate further expansion leading to irreversible changes
(eruption) of the local magnetic field. As a consequence, at least a
part of the plasma escapes from the Sun.


\begin{table}[!t]
\caption{Morphological resemblance between XPEs and EUV ejections}
\label{tab:tab3}
\begin{center}

\begin{tabular}{@{\hspace{0.5cm}}l@{\hspace{0.5cm}}l@{\hspace{0.5cm}}l}
\hline
\noalign{\hrule height 1.2 pt}

\multicolumn{3}{r}{\vspace{0.07 cm}EUV EJECTIONS} \\
\cline{2-3}
XPEs &collimated&loop-like \\

\hline
\noalign{\hrule height 1.2  pt}
collimated (\textbf{16}) &  \textbf{9}& 7\\

\hline
loop-like\hspace{0.3 cm}  (\textbf{42}) &3 &\textbf{39} \\

\hline
\end{tabular}
\end{center}
\end{table}

\begin{table}[t]
\caption{Kinematical resemblance between XPEs and EUV ejections}
\label{tab:tab4}
\begin{center}

\begin{tabular}{l@{\hspace{1.0cm}}l@{\hspace{1.0cm}}l}
\hline
\noalign{\hrule height 1.2 pt}

\multicolumn{3}{r}{\vspace{0.07 cm}EUV EJECTIONS} \\
\cline{2-3} 
XPEs &  confined& eruptive \\

\hline
\noalign{\hrule height 1.2 pt}
confined (\textbf{28}) &  \textbf{21}& 7\\

\hline
eruptive  (\textbf{30}) &5 &\textbf{25} \\

\hline
\end{tabular}
\end{center}
\end{table}


In the third criterion multiplicity of the occurrence  is considered, therefore unique XPEs that occurred once (single) are separated from recurrent events for which repeated expanding structures can be seen with time (recurrent).

In this work, we categorized EUV ejections observed by {\it TRACE} according to the classification scheme developed in the XPEs catalog. We considered similarities between the associated XPEs and the ejections observed by {\it TRACE} for each criterion separately.

Table \ref{tab:tab3} summarizes the results of our investigation for morphological criterion. In the first column we put the names of subclasses for morphological criterion and in the brackets are the total numbers of XPEs associated with eruptions seen in the EUV by {\it TRACE}. In two other columns quantity of associated TEs, is given. The bold numbers situated on the diagonal show morphological resemblance between XPEs and EUV ejections. The similarity between these subclasses is clearly seen, especially for loop-like events (92.9\%).

Table \ref{tab:tab4} summarizes the obtained results for the kinematical criterion. It is organized in the same way as Table \ref{tab:tab3}. Here we can note that the XPEs associated with EUV ejections indicated kinematical similarities, 85\% and 83\% for confined and eruptive events, respectively.

 In Table \ref{tab:tab5} the results of congruity multiplicity of occurrence are summarized.
In this case, we also point out the resemblance between the associated XPEs and the EUV eruptions.
Almost 71\% recurrent {\it TRACE} ejections  are correlated with recurrent XPEs and slightly less than 80\% single EUV eruptions show close interrelation with XPEs from the same subclass. 

\begin{table}[!h]
\caption{Recurrence resemblance between XPEs and EUV ejections}
\label{tab:tab5}
\begin{center}

\begin{tabular}{l@{\hspace{1.2cm}}l@{\hspace{1.2cm}}l}
\hline
\noalign{\hrule height 1.2pt}

\multicolumn{3}{r}{\vspace{0.07 cm}EUV EJECTIONS} \\
\cline{2-3} 
XPEs &  single& recurrent \\

\hline
\noalign{\hrule height 1pt}
single \hspace{0.58 cm}(\textbf{34}) &  \textbf{27}& 7\\

\hline
recurrent  (\textbf{24}) &7 &\textbf{17} \\

\hline
\end{tabular}
\end{center}
\end{table}


\section{Evolution of individual events}
We selected two examples of eruptions observed in EUV and soft X-ray wavelengths. For these we measured the positions of expanding features and analyzed their kinematical evolution.
\subsection{25 July 2000}
The XPE occurred in NOAA 9097 (N$06^{\circ}$, W$08^{\circ}$) at 02:47 UT on 25 July 2000. This event was associated with soft and hard X-ray flares a CME, and an EUV ejection. 

We classified this event as a loop-like, eruptive, single event. The EUV eruption was also categorized in the same way as a loop-like, eruptive, single.

\begin{figure}[!t]

\epsfig{file=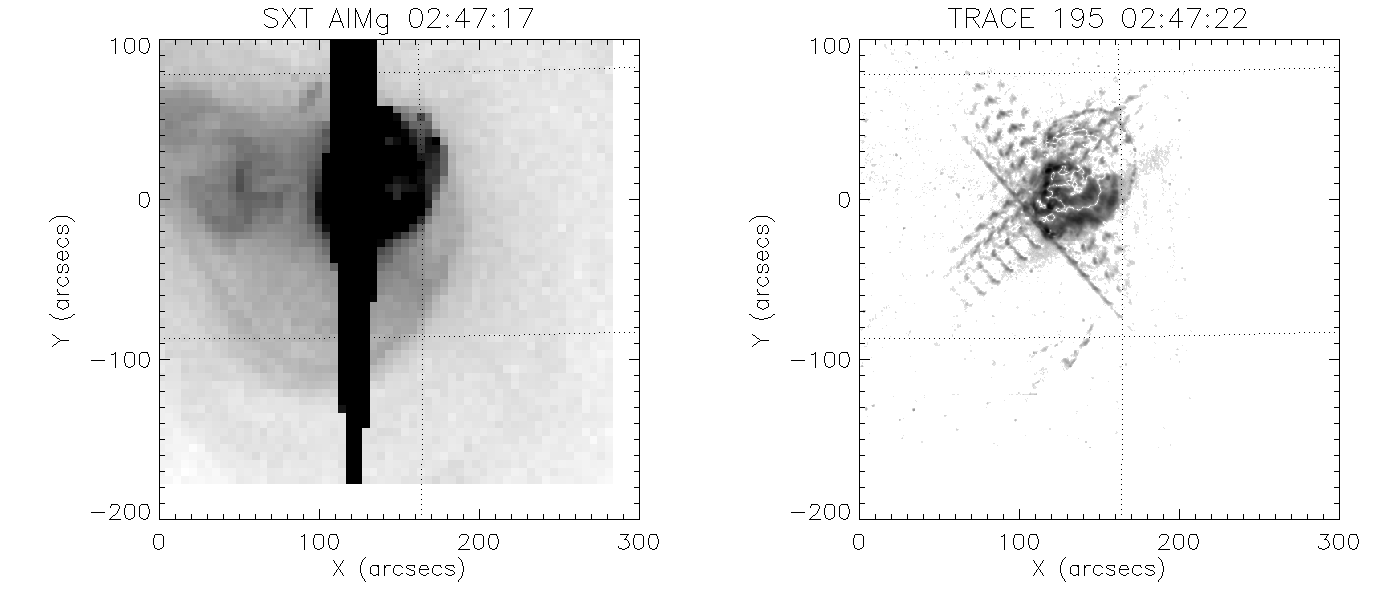,width=9.5 cm} 
\epsfig{file=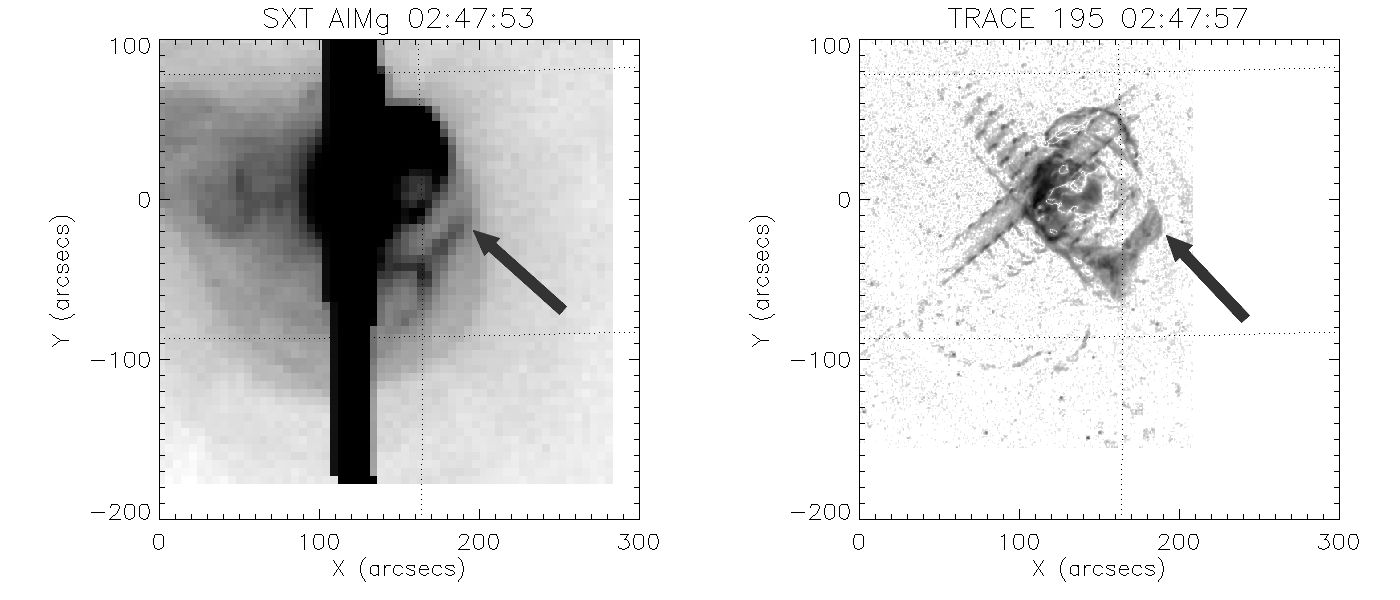,width=9.5 cm} 
\epsfig{file=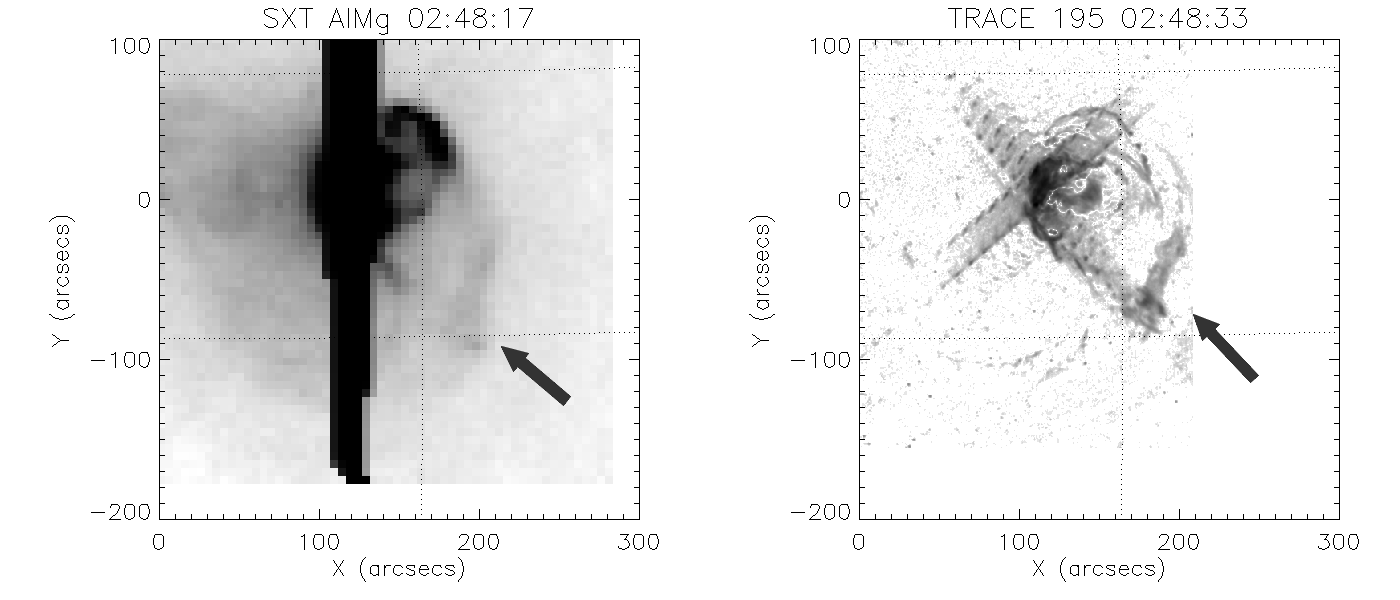,width=9.5 cm} 

\caption{A series of images of a plasmoid ejection on 2000 July 25 in two different wavelength ranges. \textit{Left}: Half-resolution images of the sandwich (AlMg) filters made in flare-mode taken from SXT.  \textit{Right}: \textit{TRACE} 195 \AA \hspace{2pt} images.Black arrows indicate the expanding front of the XPE. Artificial effects such as heavy saturation and diffraction pattern for SXT and {\it TRACE} images were seen. }
\label{fig:fig1}
\end{figure}

The EUV ejection was observed in several different passbands (1600, 171, 195 \AA). 
Figure \ref{fig:fig1} shows a set of images taken from {\it Yohkoh}/SXT and {\it TRACE} 195 \AA\hspace{0.1pt} covering the same field of view, for different times. 
The sequence of snapshot images reveals the morphological similarity between eruptions observed in the EUV and soft X-rays.The expanding loop is clearly seen in images taken in these wavelengths. 
The start of the eruption observed in X-rays and the EUV occurred almost simultaneously.

\begin{figure}[t]

\epsfig{file=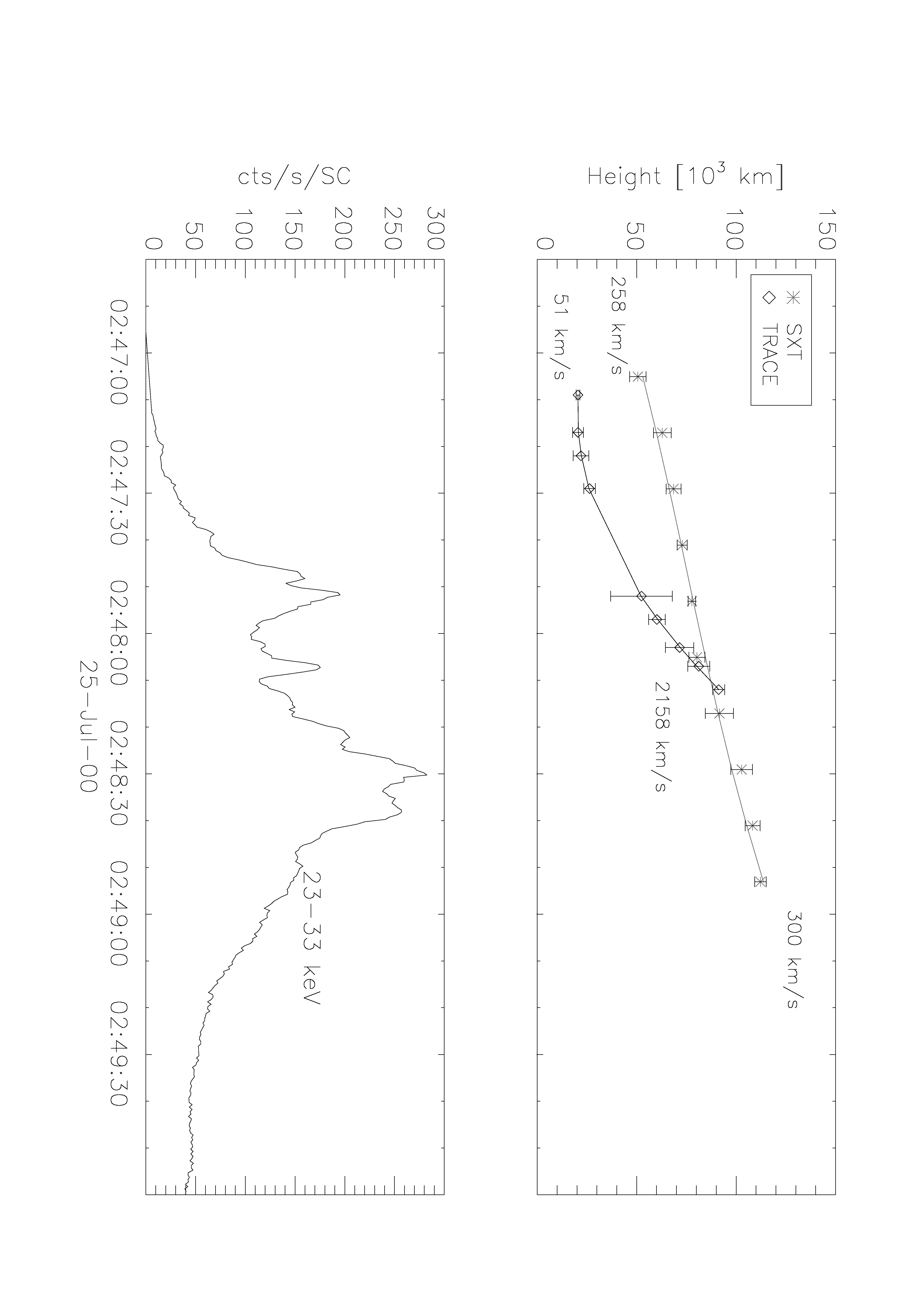,width=10 cm,angle=90} 
\vspace{-40pt}
\caption{\textit{Top}: Time profiles of the leading front of eruption height, observed in EUV and X-rays by \textit{TRACE} and \textit{Yokoh}/SXT, respectively. Apparent height was measured above some reference point. Error bars represent upper and lower limits for estimates of the plasmoid location. The apparent initial and final velocities are also shown. \textit{Bottom}: The hard X-ray flux in the energy range of $23-33$ keV is from \textit{Yohkoh}/HXT}
\label{fig:fig2}
\end{figure}

An apparent displacement of the XPE and the EUV ejection above the same reference point as a function of time for these two instruments is presented in Fig. \ref{fig:fig2}. Error bars are calculated by measuring the position of the front eruption in several different locations. 
From the height-time profiles we found that the EUV ejection showed a strong acceleration lasting less than 1.5 minutes followed by the propagation phase.
This impulsive acceleration phase of the EUV ejection took place just before the impulsive phase in hard X-rays. 
We also found that the initial and final velocities of the XPE were 258 km s$^{-1}$ and 300 km s$^{-1}$, respectively.
The values of the velocity are the apparent velocities, without considering the projection effect.


\begin{figure}[!t]
\epsfig{file=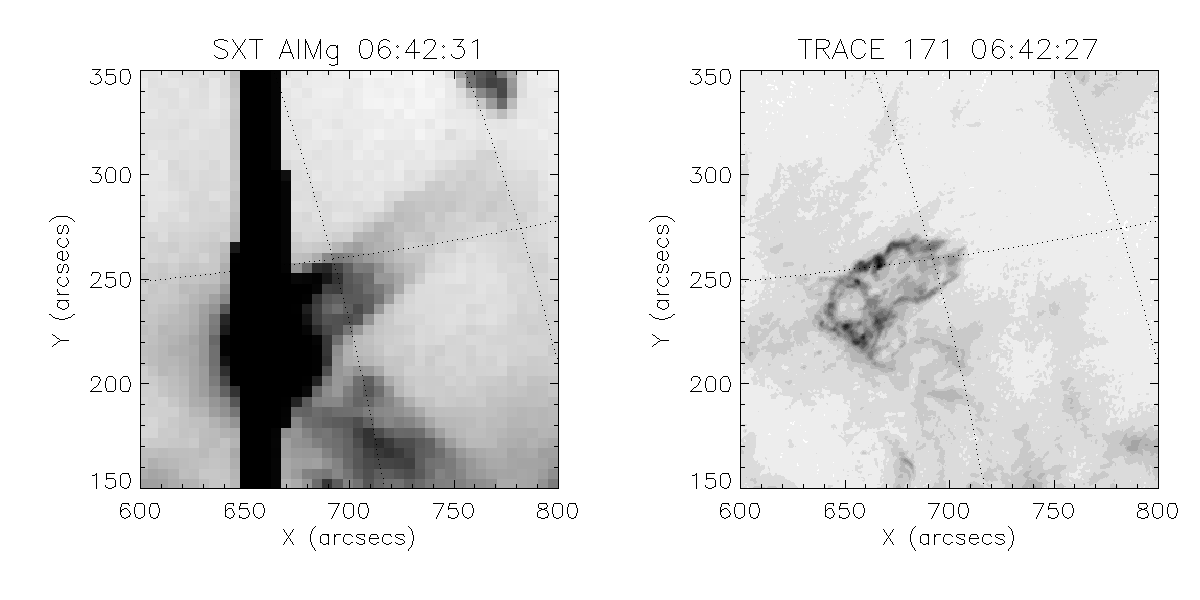,width=9.5 cm} 
\epsfig{file=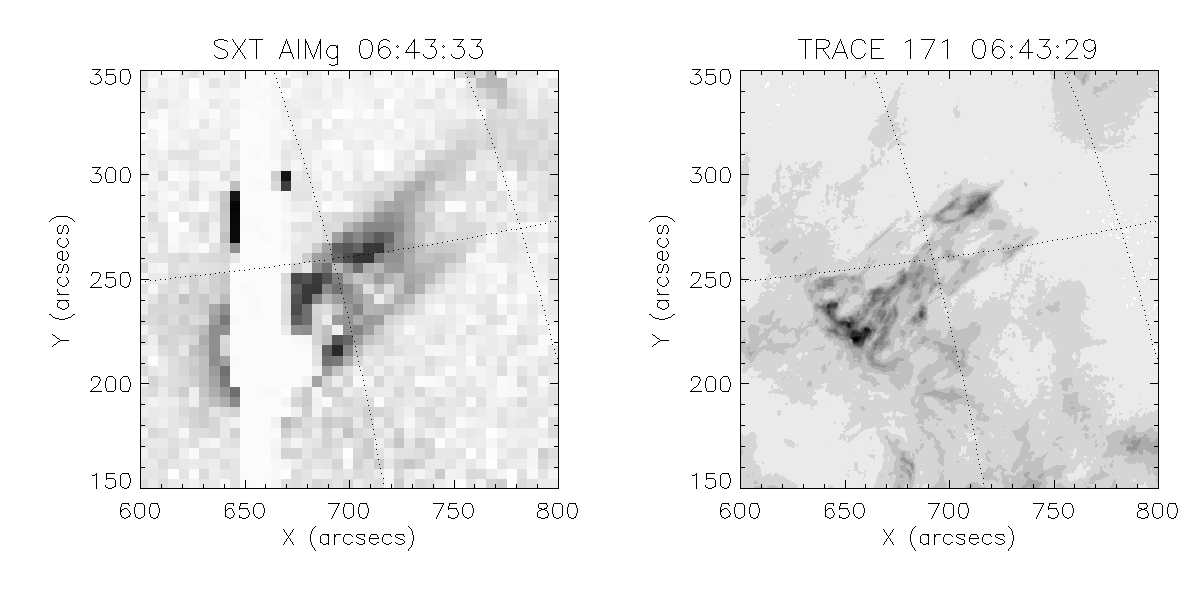,width=9.5 cm} 
\epsfig{file=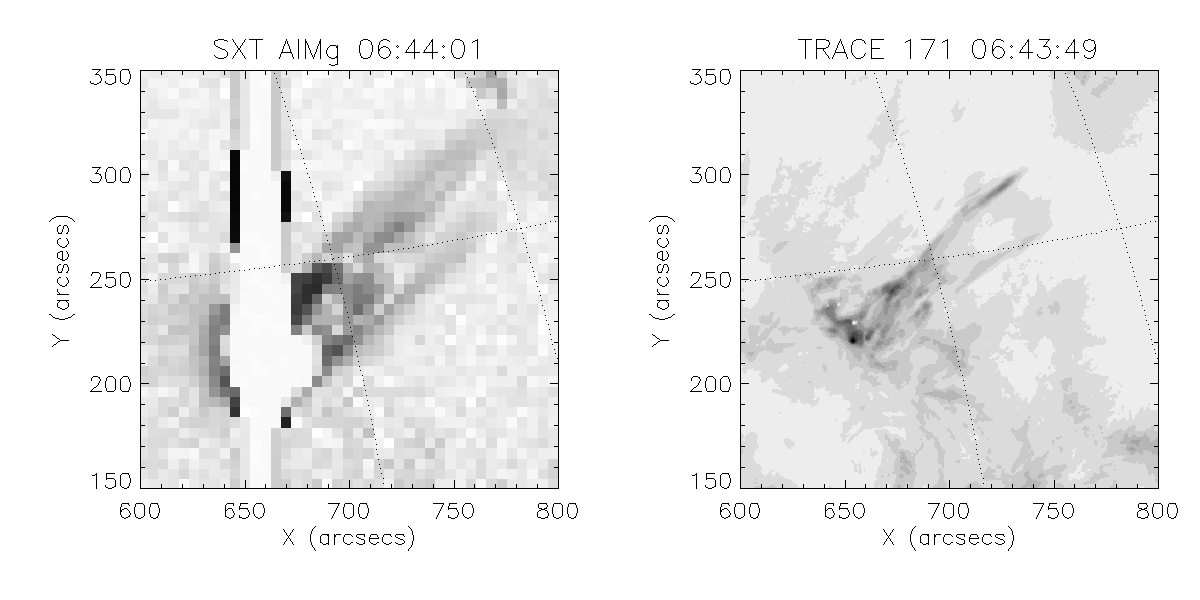,width=9.5 cm} 
\vspace{-20pt}
\caption{A series of images of plasmoid ejection in two different wavelength ranges, that occurred on 2001 October 3. \textit{Left}: Half-resolution images of the sandwich (AlMg) filters made in flare-mode taken from SXT.  \textit{Right}: \textit{TRACE} 171 \AA \hspace{2pt} images. Artificial effects such as heavy saturation and diffraction pattern for SXT and {\it TRACE} images were seen }
\label{fig:fig4}
\end{figure} 


\begin{figure}[!t]

\epsfig{file=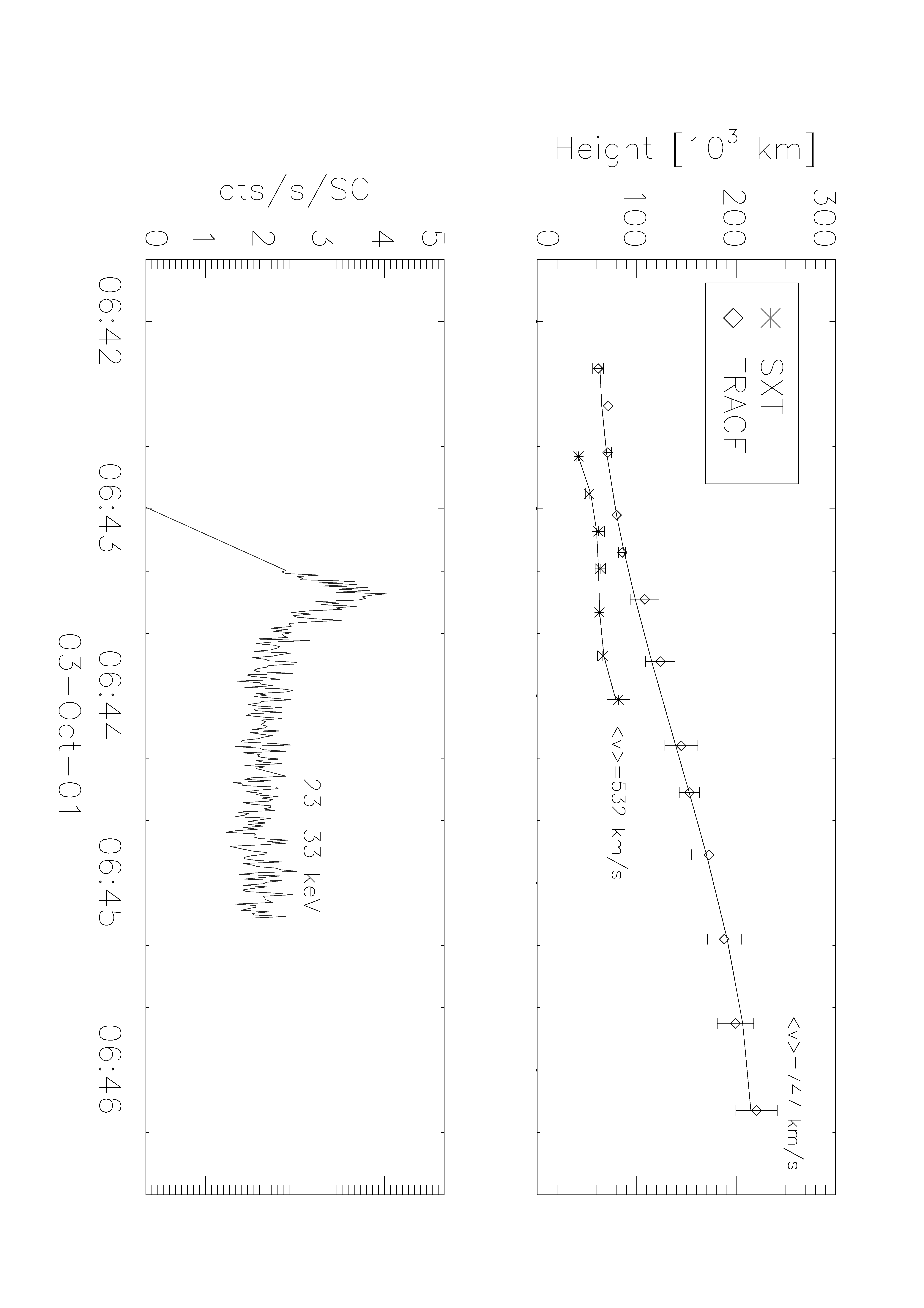,width=10 cm,angle=90} 
\vspace{-20pt}
\caption{\textit{Top}: Height-time evolution of the leading front of eruption observed in EUV and X-ray by \textit{TRACE} and \textit{Yokoh}/SXT, respectively. Apparent height was measured above some reference point. Mean values of velocity are also shown. \textit{Bottom}: The hard X-ray flux in the energy range of $23-33$ keV is from \textit{Yohkoh}/HXT}
\label{fig:fig3}
\end{figure}


\newpage

\newpage

\subsection{3 October 2001}
Our second eruption occurred in NOAA 9636 (N$19^{\circ}$, W$46^{\circ}$) on 3 October 2001 at 06:42 UT. 
This event was associated with a C6.1 flare and a CME.

This phenomenon was well observed by {\it TRACE} (171 \AA)\hspace{1 pt} and SXT (Fig. \ref{fig:fig4}). In the SXT and {\it TRACE} 171 \AA\hspace{1 pt} images, we observed a collimated, eruptive, and recurrent ejection. 
Despite of these similarities, the start of the EUV ejection was observed about 40 seconds before the start of the XPE.

From the height-time profiles (Fig. \ref{fig:fig3}) we obtained a mean velocity of 747 km s$^{-1}$ for the EUV ejection and 532 km s$^{-1}$ for the XPE.
Just before the main peak of the hard X-ray emission, warm eruption observed by {\it TRACE} was accelerated and after that the propagation phase followed.
The height-time curve for the XPE shows that the impulsive acceleration phase started before the main peak of the hard X-ray emission, as well.

\section{Conclusions}
In this paper we reported the relationship between hot and warm plasma ejetions observed in the solar corona.
Our analysis shows that the examined XPEs and the EUV ejections are strongly correlated about 90\%, even better than 
the XPEs and CMEs or the prominences and CMEs, the correlation in these last two cases being about 70\%.
Keeping in mind that XPEs are poorly associated with prominences (about 30\%),  we conclude that warm and hot plasma ejections are most closely related.

Morphological and kinematical properties as well as recurrent EUV eruptions show strong resemblance to the XPEs. 
We obtained compliance in about 83\%, 79\%, 76\% for morphological, kinematical and recurrence criteria, respectively.

We also investigated the relative locations of hot and warm plasma ejections. 
For 40 samples we found 20 XPEs that reached greater heights, and 20 EUV ejetions that were observed higher than the XPEs.
The results of this analysis may suggest that the relative location of hot or warm plasma is dependent on the individual conditions
and configurations of the magnetic field in active regions.

Detailed analysis of selected phenomena also reveals similar nature of XPEs and the EUV ejections.
Almost for all cases, we noticed the same kinematical scenario described by Zhang {et al.}, 2001- impulsive acceleration phase of these warm and hot ejections took place during impulsive phase in hard X-rays.

In the future, we plan to analyze some interesting and well-observed events recorded by the Atmospheric Imaging Assembly (AIA) on board the {\it Solar Dynamics Observatory} ({\it SDO}).
This first observatory of the NASA's Living With a Star Program gives us unusual opportunities to examine eruptive phenomena because of multiple simultaneous, high-spatial resolution,
full-disk images of the corona and transition region up to 0.5 R$_{\odot}$ above the solar
limb with unprecedented cadence (12 s) (Lemen {it et al.}, 2012).

\section*{Acknowledgements}
The {\sl Yohkoh} satellite was a project of the Institute of Space
and Astronautical Science of Japan. {\it TRACE} was a project of NASA. We acknowledge financial support from the Polish National Science Centre grants: 2011/03/B/ST9/00104, 2011/01/M/ST9/06096.

\newpage

\section*{References}
\begin{itemize}
\small
\itemsep -2pt
\itemindent -20pt

\item[]Chmielewska, E., \&  Tomczak, M. 2012, 
{\it Cent. Eur. Astrophys. Bull.}, {\bf 36}, 41

\item[] Gilbert, H.R., Holzer, T.E., Burkepile, J.T., Hundhausen, A. J.: 2000,
 {\it Astrophys. \, J.}, {\bf 537}, 503.

\item[] Gopalswamy, N., Shimojo, M., Lu, W., Yashiro, S., Shibasaki, K., Howard, R.A.: 2003,
{\it Astrophys.\, J.}, {\bf 586}, 562.

\item[] Kim,\,Y.-H., Moon,\,Y.-J., Cho,\,K.-S., {\it et al.}: 2005a,
{\it Astrophys.\,J.}, {\bf 622}, 1240.

\item[] Kim,\,Y.-H., Moon,\,Y.-J., Cho,\,K.-S., {\it et al.}: 2005b,
{\it Astrophys.\,J.}, {\bf 635}, 1291.

\item[]Lemen, J. R., Title, A M., Akin, D.J.,  Paul F. Boerner, P.F., Chou C. {\it et al.}: 2012 
{\it Solar Phys}, {\bf 275}, 17.

\item[]Ohyama, M., \&  Shibata, K. 1998,
{\it Astrophys.\,J.}, {\bf 499}, 934. 

\item[]Ohyama, M., \&  Shibata, K.: 2008, 
{\it Publ.Astron. Soc. Japan}, {\it 60}, 85.

\item[] Tomczak, M. 2003,
in Proc. ISCS 2003 Symp., Solar Variability as an Input to the Earth's Environment, ed. A. Wilson, (ESA SP-535; Noordwijk: ESA), {\bf} 465.

\item[]Tomczak, M. 2004, 
{\it A\&A},{\bf 417}, 1133.

\item[]Tomczak, M., Chmielewska, E..: 2012,
{\it Astrophys. J. Suppl.}, {\bf 199}, 10.

\item[]Tomczak, M., \&  Ronowicz, P. 2007, 
{\it Cent. Eur. Astrophys. Bull.}, {\bf 31}, 115.

\item[]Zhang J., Dere K., Howard R.A.,{\it et al.}: 2001
{\it Astrophys.\,J.}, {\bf 559}, 452.

\end{itemize}
\bibliographystyle{ceab}


\end{document}